# Atomic spin chain realization of a model for quantum criticality

R. Toskovic[1†], R. van den Berg[2†], A. Spinelli[1], I. S. Eliens[2], B. van den Toorn[1], B. Bryant[1], J.-S. Caux[2] and A. F. Otte[1]*

[1]Department of Quantum Nanoscience, Kavli Institute of Nanoscience, Delft University of Technology, Lorentzweg 1, 2628 CJ Delft, The Netherlands

[2]Institute for Theoretical Physics, University of Amsterdam, Science Park 904, 1090 GL Amsterdam, The Netherlands

[†] These authors contributed equally.
* a.f.otte@tudelft.nl

**The ability to manipulate single atoms has opened up the door to constructing interesting and useful quantum structures from the ground up[1]. On the one hand, nanoscale arrangements of magnetic atoms are at the heart of future quantum computing and spintronic devices[2,3]; on the other hand, they can be used as fundamental building blocks for the realization of textbook many-body quantum models, illustrating key concepts such as quantum phase transitions, topological order or frustration[4]. Step-by-step assembly promises an interesting handle on the emergence of quantum collective behavior as one goes from one, to few, to many constituents. To achieve this, one must however maintain the ability to tune and measure local properties as the system size increases. Here, we use low-temperature scanning tunneling microscopy to construct arrays of magnetic atoms on a surface, designed to behave like spin-1/2 XXZ Heisenberg chains in a transverse field, for which a quantum phase transition from an antiferromagnetic to a paramagnetic phase is predicted in the thermodynamic limit[5]. Site-resolved measurements on these finite size realizations reveal a number of sudden ground state changes when the field approaches the critical value, each corresponding to a new domain wall entering the chains. We observe that these state crossings become closer for longer chains, indicating the onset of critical behavior. Our results present opportunities for further studies on quantum behavior of many-body systems, as a function of their size and structural complexity.**

Since the birth of quantum mechanics, lattice spin systems[6] have represented a natural starting point for understanding collective quantum dynamics. Today, scanning tunneling microscopy (STM) techniques allow to experimentally build and probe realizations of exchange-coupled lattice spins in different geometries[7,8]. In linear arrangements, quantum effects are strongest[9] and notions such as quantum phase transitions[10] are most easily understood, the simplest illustration being the Ising model in a transverse field[11,12]. In this work, using STM, we construct finite-size versions of a model in the same universality class, namely the spin-1/2 XXZ chain in a transverse field[5], which has previously been realized in the bulk material $Cs_2CoCl_4$ [13,14]. Our setup allows us to probe the chains with single spin resolution while tuning an externally applied transverse field through the critical regime.

The chains are created by evaporating Co atoms onto a $Cu_2N/Cu(100)$ surface (see Supplementary Information), which provides efficient decoupling for the magnetic d-shell electrons from the underlying bulk electrons[7]. Employing inelastic electron tunneling spectroscopy (IETS)[15,16] at sufficiently low temperature (330 mK) allows us to determine the magnetic anisotropy vector of each atom[17] as well as the strength of the exchange coupling between neighboring atoms[18]. It was previously demonstrated that Co atoms on this surface behave as spin $S = 3/2$ objects experiencing a strong uniaxial hard-axis anisotropy pointing in-plane, perpendicular to the bond with the neighboring N atoms[19]. As a result, the $m_z = \pm 3/2$ states split off approximately 5.5 meV above the $m_z = \pm 1/2$ doublet (see Fig. 1a). As we will show below, by exploiting the magneto-crystalline anisotropy, we thus effectively reduce the spins from 3/2 to 1/2. The $Cu_2N$ islands were kept small (~ 6 nm) in order to ensure limited variation in anisotropy and substrate coupling between different atoms inside the chains[20].

The Co atoms are manipulated into the arrangement shown in Fig. 1b, such that their interaction is governed by the spin 3/2 nearest neighbor antiferromagnetic isotropic Heisenberg exchange:

$$\mathcal{H}_{3/2} = J \sum_{i=1}^{N-1} \mathbf{S}_i \cdot \mathbf{S}_{i+1} + D \sum_{i=1}^{N} (S_i^z)^2 - g\mu_B B_x \sum_{i=1}^{N} S_i^x ,$$

with interaction strength $J = 0.24$ meV [21], subjected to an external magnetic field **B** (with g-factor $g = 2.3$ [19]) applied perpendicular to the surface. Since all other relevant energy scales ($k_B T$, $\mu_B B$) stay well below the anisotropy energy $2D \approx 5.5$ meV, excitations to $\pm 3/2$ doublets can be projected out through a Schrieffer-Wolff transformation up to first order in $1/D$ [14,22,23]. This results in an effective spin-1/2 Hamiltonian:

$$\mathcal{H}_{1/2} = \sum_{i=1}^{N-1} J_\perp \left( S_i^x S_{i+1}^x + S_i^y S_{i+1}^y \right) + J_z S_i^z S_{i+1}^z + J_\perp^{\mathrm{nnn}} \sum_{i=1}^{N-2} S_i^x S_{i+2}^x + S_i^y S_{i+2}^y - \mu_B B_x \sum_{i=1}^{N} g_i S_i^x ,$$

with nearest and next-nearest neighbor exchange parameters and bulk/boundary g-factors given by:

$$J_\perp = 4J, \quad J_z = J - \frac{39 J^2}{8D}, \quad J_\perp^{\mathrm{nnn}} = -\frac{3 J^2}{D}, \quad g_i = \begin{cases} 2g \left(1 - \frac{3J}{2D}\right) & \text{if } i = 2, ..., N-1 \\ 2g \left(1 - \frac{3J}{4D}\right) & \text{if } i = 1, N \end{cases} .$$

The next-nearest neighbor coupling $J_\perp^{\mathrm{nnn}}$ generated by the Schrieffer-Wolff transformation does not affect the qualitative features of the spectrum nor the existence of the phase transition, effectively reducing $H_{1/2}$ to an XXZ Hamiltonian in a transverse field.

Figs. 1c—d show the calculated lowest excitation energies of $H_{1/2}$ for an even- ($N = 8$, panel c) and an odd-numbered ($N = 9$, panel d) chain, for a transverse field up to 9 T. Below the transition to the paramagnetic phase, just below 6 T, several ground state crossings are predicted and their number increases with chain length. Starting from a state with Néel-like order near zero field, each crossing corresponds to a stepwise increase of the total magnetization $M$ along the field and the average number of antiferromagnetic domain walls $n$ inside the chain.

The lowest excited state is energetically distinguishable in finite chains, but becomes degenerate with the ground state in the thermodynamic limit, where it corresponds (via a Jordan-Wigner transformation[24,25]) to the topological edge-states recently observed in ferromagnetic chains on a superconducting surface[26]. Below the critical field $B_{\mathrm{crit}}$, the ground state and this zero mode are separated from the higher excited states by an energy gap $E_G$. As the length of the chain increases, $E_G$ remains finite and forms the characteristic energy separating the ground state from the continuum, except at $B_{\mathrm{crit}}$, where it vanishes. Just below this point, spin liquid behavior is predicted[5].

We constructed chains of Co atoms of various length and performed low-temperature IETS measurements ($T = 330$ mK $< E_G/k_B$) on each atom in a chain while varying the strength of the transverse field. To obtain an extensive dataset, a fully automated measurement sequence was employed (footnote). Figs. 2a—b show measurements taken on the first atom of an odd-length (5 atoms, panel a) and an even-length chain (6 atoms, panel b), recorded for every 200 mT from 0 to 9 T. At voltages below 5.5 mV, transitions within the manifold of $m_z = \pm 1/2$ states are observed; excitations at higher voltages correspond to transitions to the $m_z = \pm 3/2$ manifold. The spectra show sudden changes in both excitation energy and intensity at field values corresponding to expected ground state crossings: near 3.5 T for $N = 5$ and near 1.5 T and 4.0 T for $N = 6$.

To simulate the differential conductance signal, we employed a perturbative transport model[16,21,27,28]. Steps related to the spectrum are found in good agreement with the data using the $S = 3/2$ Hamiltonian (eq. 1), Figs. 2c—d. Calculations using the $S = 1/2$ XXZ Hamiltonian (eq. 2, Figs. 2e—f) show similar agreement, except for the excitations to the $m_z = \pm 3/2$ multiplet near $\pm 5.5$ mV, which are not modeled. This agreement justifies our effective spin-1/2 treatment. Notable quantitative discrepancy between theory and experiment is found near 1.5 T in the $N = 6$ chain. At this field value, a twofold ground state degeneracy occurs, resulting in a zero bias Kondo resonance in the data, which is only partly reproduced in the third order perturbative analysis[21,28–31].

In Fig. 3, field dependent measurements are shown for all atoms of chains of 1 to 9 atoms, featuring a total of 2056 IETS spectra. Here, we focus on the ±3 mV range corresponding to the $m_z = \pm 1/2$ multiplets. As chain length increases, more features become visible. When comparing these to the calculated ground state crossing positions (lower panels), we find that for chains up to length 6 each feature lines up with one of the crossings.

For longer chains the agreement becomes less good, though qualitatively the observed data evolve as expected. Specifically, for atoms in the bulk of the chain (two or more sites away from an edge), a continuous featureless region is observed between 3 T and 6 T, which widens as chain length increases. In this field range ground state crossings become too close to be individually resolved. The energy difference between the ground state and the zero mode also decreases, such that their thermal occupations become comparable. This further reduces the ability to resolve the crossings.

A simplified picture in terms of spin-1/2 product states and spin flip operations[8] provides a qualitative understanding of the IETS spectra. In even chains, for small fields the two Néel orderings are equally mixed in the ground state. Here the first crossing is predominantly found on the outer atoms since at this crossing the number of domain walls $n$ increases by 1 only. In odd length chains, the magnetic field selects one of these Néel states leading to a definite staggered magnetization profile: flipping the spin of an odd (even) atom points it against (along) the magnetic field, leading to a state increasing (decreasing) in energy with increasing field. At fields above the critical field, the ground state is essentially polarized and we can obtain a similar understanding in terms of magnon physics.

The semiclassical reasoning outlined above is further confirmed by measurements taken on a 7-atom chain with a spin-polarized (SP) STM tip, shown in Fig. 4. In contrast to SP-STM measurements taken at a fixed voltage[8], these spectra reveal spin contrast in energy dependent phenomena such as spin excitations. At 3 T we see, in addition to the even-odd pattern in the excitation energies, an alternating pattern in spin excitation intensities (Fig. 4c). For positive sample bias, in which case an excess of spin down electrons from the tip is injected into the chain (Fig. S1), excitations on odd-site spins are enhanced. At negative voltage, excitations are enhanced on the even sites. This alternating pattern is found to disappear as the field is swept through the critical value (Fig. 4d). Additional SP-STM data are shown in Fig. S2.

In conclusion, we have built chains of effective $S = 1/2$ spins realizing the XXZ model in a transverse field, and obtained detailed site-resolved information about the spectrum as a function of chain length and applied field. Increasing the chain length shows a growing number of ground state crossings, a precursor of the Ising quantum phase transition occurring in the thermodynamic limit. The origin of the discrepancy between the theoretical positions of ground state crossings and the ones observed in longer chains remains an open issue that requires a better understanding of the electronic and magnetic structure of the chains and their supporting surface. Our work furthermore shows that the Co chains form Kondo lattices; the exact description of the Kondo resonances remains a theoretical challenge.

This work was supported by FOM, NWO, the Delta ITP consortium and by the Kavli Foundation. We thank M. Ternes for developing the third order perturbative transport model.

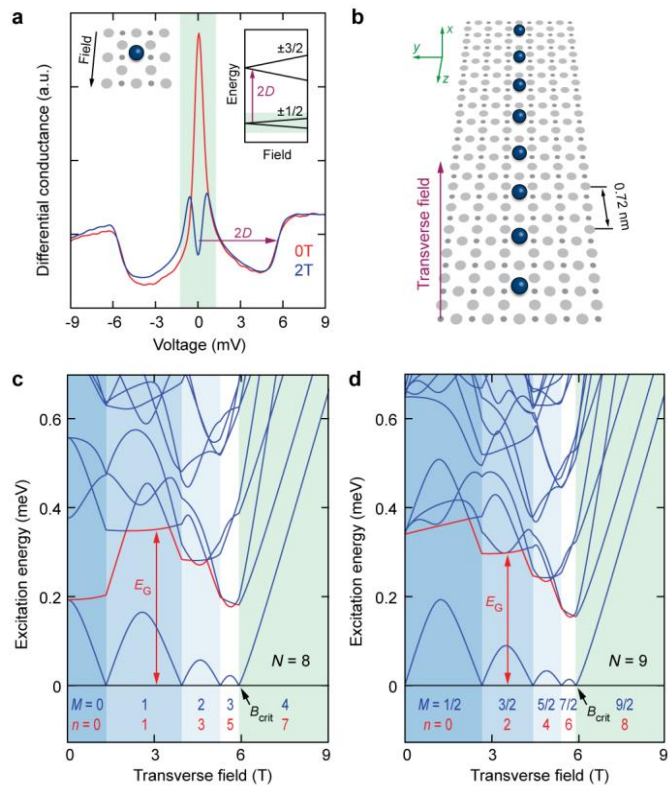

**Fig. 1. Construction of XXZ chains. a.** IETS spectra taken on a single Co atom on $Cu_2N$ at 0 T and 2 T applied along the hard axis. Left inset: atomic arrangement near the Co atom. Right inset: energy diagram indicating the separation between the ±1/2 and ±3/2 doublets. **b.** Atomic design for XXZ chains and indication of the transverse field direction. Large (small) grey circles represent Cu (N) atoms. **c.** Lowest excitation energies of an $N = 8$ chain for a transverse field up to 9 T. $E_G$ and $B_{crit}$ are indicated, as well as transverse magnetization $M$ and average number of domain walls $n$ between each ground state change. **d.** Same as (c) for $N = 9$.

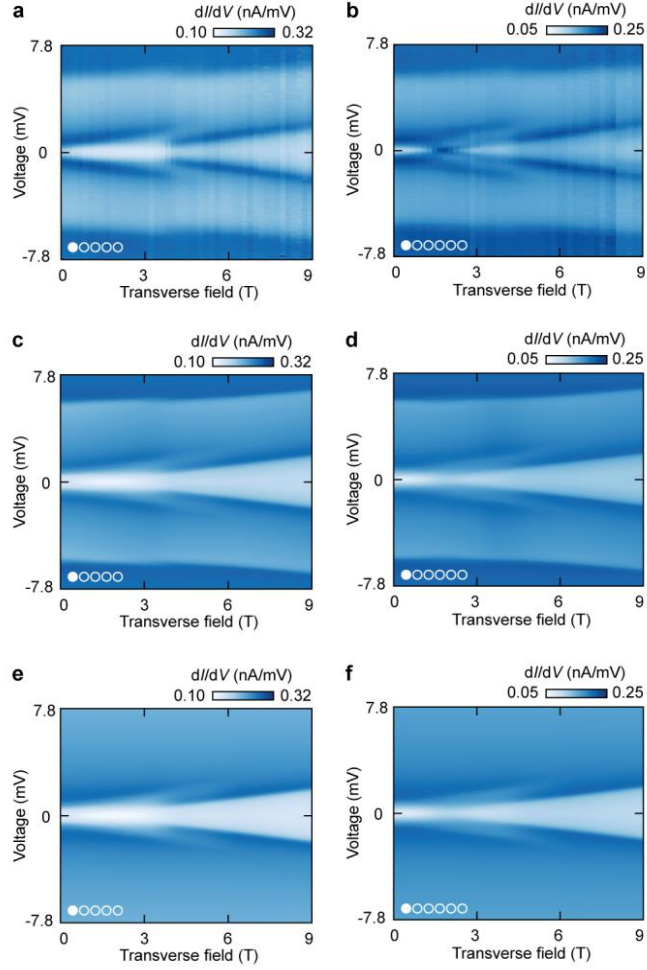

**Fig. 2. Comparison to theory. a.** IETS spectra taken on atom 1 of an $N = 5$ chain in transverse fields ranging from 0 T to 9 T, in increments of 200 mT. **b.** Same as (a), but taken on atom 1 of an $N = 6$ chain. IETS curves were normalized in order to correct for tip height variations. Conductance values listed at the color bars are indicative only: due to normalization, scaling between spectra may vary by ~20%. **c—d.** Theoretical spectra corresponding to (a—b), calculated using a spin-3/2 model (eq. 1). The Kondo peak appearing at the first ground state crossing in (b) is underrepresented in the theory (d). **e—f.** Same as (c—d), but calculated using a spin-1/2 XXZ model (eq. 2).

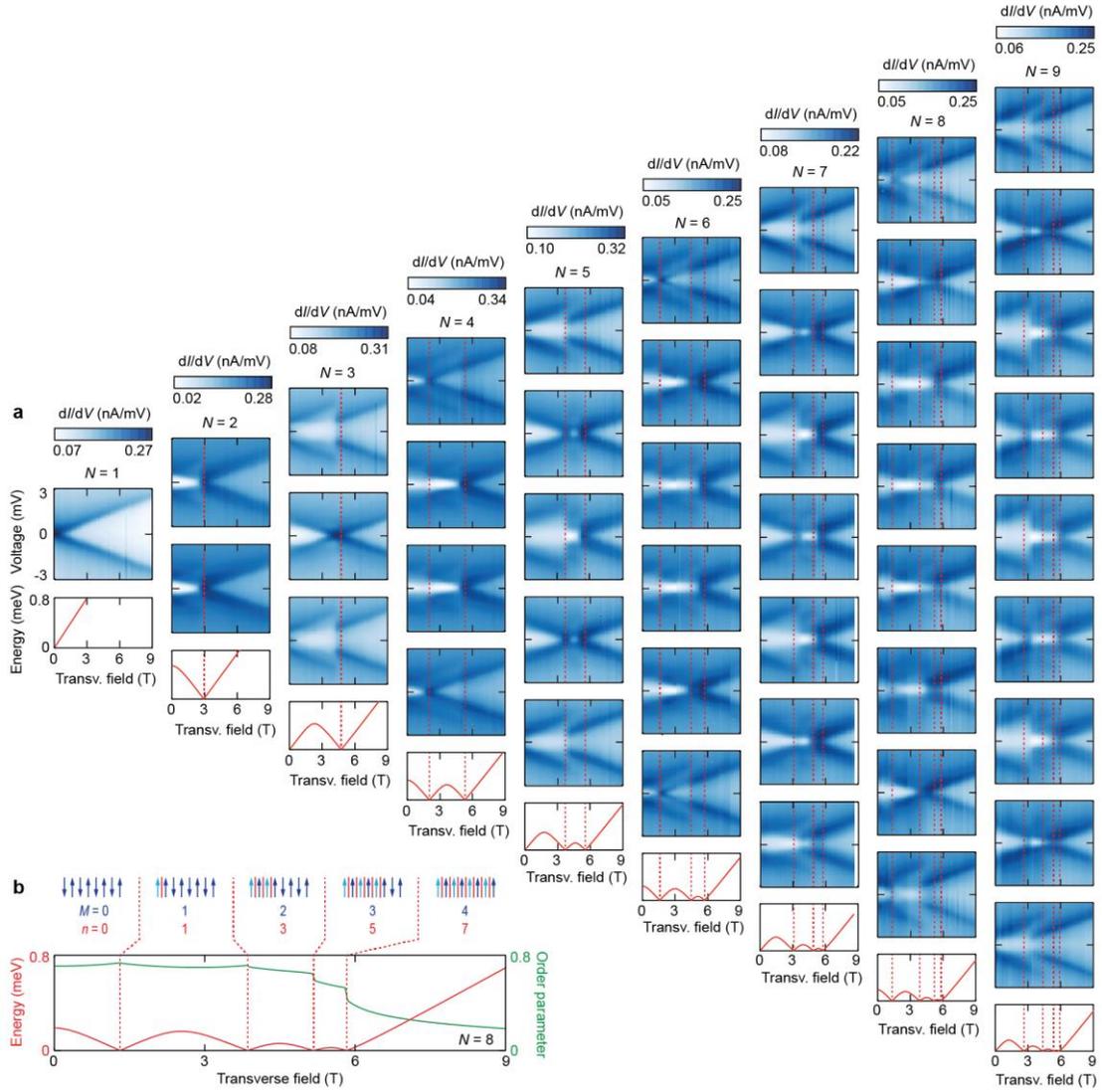

**Fig. 3. Experimental results on chains of 1 to 9 atoms. a.** ±3 mV IETS spectra from 0 T to 9 T transverse field (in 200 mT increments) obtained on each atom of every chain up to a length of 9 atoms (up to 8.6 T for $N = 7$). Red dashed lines indicate positions of expected calculated ground state crossings. Due to normalization, scaling of individual spectra may differ by ~20% from values listed at the color bars. **b.** Calculated staggered order in $y$ direction for an 8-atom chain (see Supplementary Information). Arrows at the top, representing magnetization along the transverse field direction ($x$), indicate typical contributions to the ground state in each region between crossings.

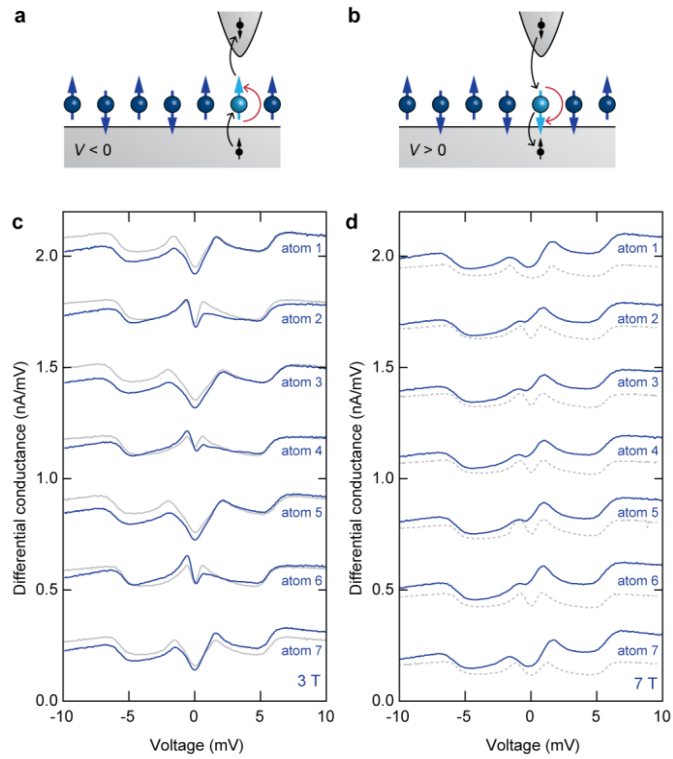

**Fig. 4. Spin-polarized spectroscopy. a—b.** Schematics showing allowed spin excitations in case of a fully polarized (in the spin-down direction, see Fig. S1) STM tip for negative and positive sample voltages. **c.** Spin-polarized IETS spectra taken at 3 T transverse field on an $N = 7$ chain (blue curves). Corresponding spectra taken with an unpolarized tip are shown in grey. **d.** Same as c, but at 7 T. Here the unpolarized data (dotted grey lines) were taken on a different, but identical chain.

**Footnote:**
Having visited each of the atoms in a chain, the STM tip returned to the first atom, after which the field was automatically increased in small increments, each followed by an atom-locking sequence. For field increments of 50 mT, the lateral drift of the tip was smaller than one atom radius, allowing the atom-locking procedure to reliably locate the atom after each increment. During field ramps the tip was raised by 2 nm. See Supplementary Information for more details.

# Supplementary information for:

# Atomic spin chain realization of a model for quantum criticality

R. Toskovic[1], R. van den Berg[2], A. Spinelli[1], I. S. Eliens[2], B. van den Toorn[1], B. Bryant[1], J.-S. Caux[2] and A. F. Otte[1]

[1]Department of Quantum Nanoscience, Kavli Institute of Nanoscience, Delft University of Technology, Lorentzweg 1, 2628 CJ Delft, The Netherlands

[2]Institute for Theoretical Physics, University of Amsterdam, Science Park 904, 1090 GL Amsterdam, The Netherlands

**Experimental set-up and measurements**

The measurements reported in this paper have been conducted in a commercial low-temperature STM (UNISOKU USM 1300S) at 330 mK and in ultrahigh vacuum (below $2\times10^{-10}$ mbar). A $Cu_2N$ monolayer was prepared *in situ* on a Cu(100) substrate by sputtering $N_2$ for 45 s at $1\times10^{-5}$ mbar and 500 eV followed by 1 min of annealing at 400°C, resulting typically in rectangular $Cu_2N$ islands with their most elongated direction being smaller than 10 nm (*1*). Co atoms were evaporated at approx. 1060°C onto the precooled $Cu_2N$ surface. A PtIr tip was used, which we prepared by e-beam annealing followed by indention into a bare Cu surface. Chains of Co atoms were assembled by means of vertical atom manipulation.

IETS measurements on Co atoms were realized by recording *dI/dV* spectra employing a lock-in technique with excitation voltage amplitude of 70 $\mu V_{RMS}$ at 928 Hz. Unless specified otherwise, in all measurements the applied magnetic field (up to 9 T) was oriented perpendicular to the sample surface. IETS measurements were performed at intervals of 200 mT, forming a set of 46 spectra per atom (except for the 7-atom chain, for which only 44 spectra up to 8.6 T were performed). In order to achieve a substantial reduction of the data acquisition time, an automated procedure was developed. After taking spectra on each atom at a given field, the tip returned to the first atom and retracted 2 nm followed by a 50 mT automated field sweep. Following each sweep, the tip was brought back in tunneling range (50 pA, 15 mV) upon which potential drift was corrected for through an automated atom-locking procedure. For field increments larger than 50 mT, the acquired drift would be larger than one atom radius, making atom-locking impossible. Using this method, obtaining a data set for fields ranging from 0 T to 7 T required performing the experiment continuously for 7 h on a single Co atom to 28 h on a $N = 9$ chain. Spectra taken above 7 T were obtained manually for each atom of every chain, due to tip instabilities disabling proper atom-locking when sweeping the field in that range.

**Order parameter**

The order parameter shown in Fig. 3b is the nearest neighbor correlator

$$-\frac{1}{N}\sum_{i=1}^{N-1}\langle S_i^y S_{i+1}^y \rangle$$

representing the staggered magnetization along the transverse *y* direction, which is oriented perpendicular to the direction of the magnetic field (*x*).

**Spin-polarized measurements**

A spin-filtering tip was created by attaching several Co atoms to the tip and applying a field of 3 T perpendicular to the surface. Spin-polarization was verified by performing spectroscopy on a single Co atom. As shown in Fig. S1, the relative heights of the peaks were found to be opposite to recently reported (*2*), indicating that the ultimate atom of the tip was polarized opposite to the external field due to exchange forces within the tip.

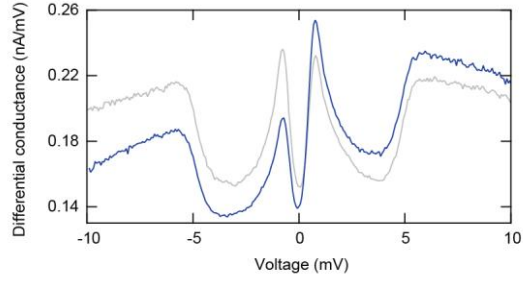

**Fig. S1.** IETS spectra on a single Co atom on $Cu_2N$ in a 3 T magnetic field oriented perpendicular to the surface, taken with the same spin-polarized tip as used in Fig. 4 (blue) and with a spin-unpolarized tip (grey).

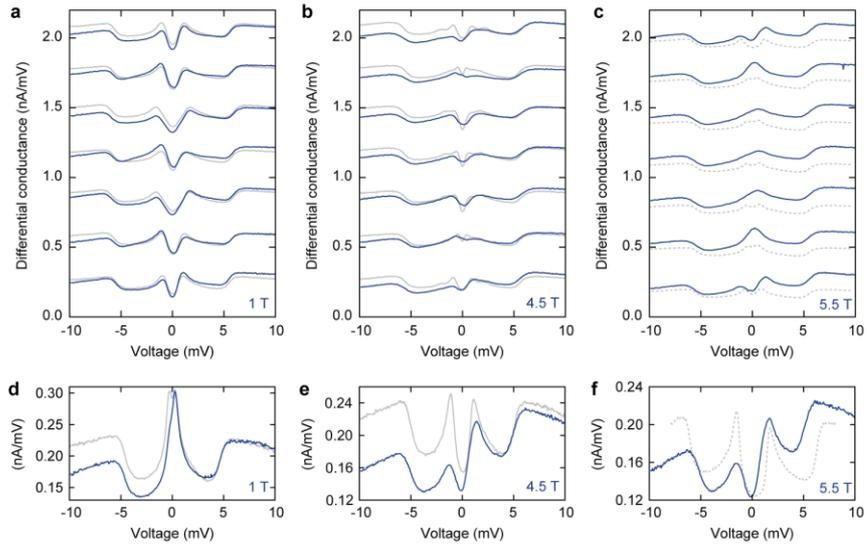

**Fig. S2. a—c.** Spin-polarized IETS spectra taken at 1 T, 4.5 T and 5.5 T transverse field in an N = 7 chain (blue curves). Corresponding spectra taken with an unpolarized tip are shown in grey. **d—e.** IETS spectra taken on a single Co atom at the same field values as in a—c with the same spin-polarized tip (blue) and with a spin-unpolarized tip (grey). Dashed spin-unpolarized curves were taken on a different instance of the chain/atom than the corresponding spin-polarized curves.